\documentclass[fleqn,12pt,twoside]{article}
\usepackage{espcrc1}

\usepackage{graphicx}
\usepackage[figuresright]{rotating}

\newcommand{\AmS}{{\protect\the\textfont2
  A\kern-.1667em\lower.5ex\hbox{M}\kern-.125emS}}

\title{Correlation femtoscopy}

\author{R. Lednick\'y\address[JINR]{Joint Institute for Nuclear
Research,\\Dubna, Moscow Region, 141980, Russia}
\address[IP]{Institute of Physics ASCR,\\
Na Slovance 2, 18221 Prague 8, Czech Republic}
\thanks{\uppercase{W}ork supported by the
Grant Agency of the Czech Republic under contract 202/04/0793
and partly carried out within the scope of the GDRE:
Heavy ions at ultrarelativistic energies -–
a European Research Group comprising
IN2P3/CNRS, EMN, University of Nantes,
Warsaw University of Technology, JINR Dubna, ITEP Moscow and
BITP Kiev.}
}

\begin{document}

\maketitle

\begin{abstract}
The basics of correlation femtoscopy, recent results
from femtoscopy in relativistic heavy ion collisions
and their consequences
are shortly reviewed.
\end{abstract}

\section{Introduction}

The momentum correlations of two or more particles at small
relative momenta in their center-of-mass (c.m.) system
are widely used to study space-time characteristics of the
production processes on a level of fm $=10^{-15}$ m,
so serving as a correlation femtoscopy tool.
Particularly, for non-interacting identical particles,
like photons or,
to some extent, pions, these correlations result from the
interference of the production amplitudes due to the
symmetrization requirement of quantum statistics
(QS) \cite{GGLP60,gkp71,KP72,kop74,pod89}.

There exists \cite{gkp71,KP72,kop74,pod89,shu73,coc74,KP75}
an analogy of the momentum QS
correlations of photons with the space--time correlations
of the intensities of classical electromagnetic fields used in
astronomy to measure the angular radii of stellar objects
based on the superposition principle (HBT effect) \cite{hbt}.
This analogy is sometimes misunderstood and the momentum
correlations are mixed up with the space-time (HBT)
correlations in spite that their orthogonal character
and thus the absence of the former in astronomy measurements
due to extremely large space-time extent of stellar objects
(and vice versa) was already pointed out in early
paper \cite{KP75}.
Note that though the space-time (HBT) correlations
are absent in the subatomic measurements, they
can still be used in laboratory
as an intensity-correlation spectroscopy
tool (see \cite{glw66,pkd67} and references therein).

The momentum QS correlations were first
observed as an enhanced production of the
pairs of identical pions with small opening angles
(GGLP effect \cite{GGLP60}). Later on,
Kopylov and Podgoretsky
settled the basics of correlation femtoscopy
in more than 20 papers (see a review \cite{pod89})
and developed it as a practical tool;
particularly,
they suggested to study the interference effect
in terms of the correlation function,
proposed the mixing techniques to construct the uncorrelated
reference sample
and clarified the role
of the space--time characteristics
of particle production in various physical situations.

The momentum correlations of particles emitted at nuclear
distances are also influenced by the effect of
final state interaction (FSI) \cite{koo,gkw79,ll1,BS86,bgj90}.
Thus the effect of the Coulomb interaction dominates the correlations
of charged particles at very small relative momenta
(of the order of the inverse Bohr radius
$|a|^{-1}$ of the two-particle system),
respectively suppressing or
enhancing the production of particles with like or unlike charges.
Though the FSI effect complicates the correlation analysis,
it is an important source of information allowing for the
coalescence femtoscopy (see, {e.g.}, \cite{sy81,lyu88,mro93,sh99}),
the correlation femtoscopy with unlike particles \cite{ll1,BS86,bgj90}
including the access to the relative space--time
asymmetries in particle production
\cite
{LLEN95,ali95,vol97,sof97,lpx,led03,mis98,led02,blu02,ada03a,kis04,led04}
and a study of strong interactions between specific particles
\cite{led02,kis04,led04}.

In this review, I will concentrate on the assumptions behind
the correlation femtoscopy formalism and discuss the recent
results obtained from the femtoscopy analysis of like and
unlike particle correlations in relativistic heavy
ion collisions.
One can inspect recent reviews \cite{led04,lis05,cso05} for
a number of important topics that are not touched here,
such as non-Gaussian tails,
imaging techniques, correlations of penetrating
probes, comparison of different colliding systems
or spin correlations.

\section{Assumptions}

The two-particle
correlation function ${\mathcal R}(p_{1},p_{2})$
is usually defined as a
ratio of the measured two-particle distribution to the
reference one obtained by mixing particles from different
events of a given class,
normalized to unity at sufficiently
large relative momenta.
The space-time information contained in the momentum
correlations is usually extracted based on the following assumptions:

(i) The mean freeze-out phase space density
$\langle f \rangle$ is assumed sufficiently
small so that only the mutual QS and FSI effects can be considered
when calculating the correlation function of two particles
emitted with a small relative momentum $Q=2 k^*$ in their
c.m. system.
This {\it two-particle approximation} may not be justified for
the rare pairs associated with large phase-space density fluctuations
and also in low energy heavy ion reactions
when the particles are produced in a mean field of residual
nuclei. To deal with this field a quantum adiabatic (factorisation)
approach \cite{3body,bar96} or the transport simulations
(see, e.g., \cite{rqmd,gon91,iss05}) can be used.
At high energies, there are also some attempts to account
for the mean field
effects in eikonal \cite{won04,zha04,kl05} and optical potential
\cite{cra05,pra05} approaches,
the latter giving some tens of percent change of the
outgoing phase space density profiles
in heavy ion collisions at RHIC assuming however unrealistic
in-medium masses at the kinetic freeze-out and neglecting
a rapid time dependence of the potential.
As for $\langle f \rangle$, it increases with energy and for
central lead-lead or gold-gold collisions seems to saturate
at the highest SPS energy \cite{lis05,as05} (see, however,
\cite{fer99}).
The saturated $\langle f \rangle$ is substantially smaller
than unity for pions with $p_t > 0.2$ Gev/$c$
so pointing to negligible multiboson effects
(see, e.g., \cite{mb00,hei00})
in this $p_t$-region.

(ii) The momentum dependence of the one-particle emission
probabilities is assumed
inessential when varying the particle four-momenta
$p_{1}$ and $p_{2}$ by the
amount characteristic for the correlations due to QS and FSI.
This {\it smoothness assumption}, requiring the components of the
mean space-time distance between particle emitters
much larger than those of the space-time extent
of the emitters, is well justified for heavy ion collisions.

(iii) An independent or incoherent particle emission is assumed.
This assumption is quite reasonable for a dominant part
of particle pairs produced in heavy ion collisions and
is consistent with the observed strength of two- and three-pion
correlation functions (see, e.g., \cite{ada03,mor03,csa05}).

(iv) To simplify the calculation of the FSI effect,
the Bethe-Salpeter amplitude describing two particles
emitted at space-time points $x_{i}=\{t_{i},{\bf r}_{i}\}$
and detected with four-momenta $p_i$
is usually calculated at equal emission times in the pair
c.m. system;
i.e. the reduced non-symmetrized Bethe-Salpeter amplitude
(with the removed
unimportant phase factor due to the c.m. motion),
depending only on the relative four-coordinate
$\Delta x\equiv x_1-x_2=\{t,{\bf r}\}$
and the generalized relative momentum
$\widetilde{q}=q-P(qP)/P^2$ ($q = p_1-p_2$,
$P=p_1+p_2$ and $qP = m_1{}^2-m_2{}^2$;
in the two-particle c.m. system, ${\bf P} = 0$,
$\tilde q = \{0,2{\bf k}^*\}$ and $\Delta x = \{t^*,{\bf r}^*\}$),
is substituted by
a stationary solution $\psi ^{S(+)}_{-{\bf k}^*}({\bf r}^*)$ of the
scattering problem having at large distances
${r}^*$ the asymptotic form of a
superposition of the plane and outgoing spherical waves
(the minus sign of the vector ${\bf k}^{*}$ corresponds to the reverse
in time direction of the emission process).
This {\it equal time} approximation is
valid on conditions \cite{ll1,led05}
$ |t^*|\ll m_{2,1}r^{*2}$ for
${\rm sign}(t^*)=\pm 1$ respectively.
These conditions are usually satisfied
for heavy particles like kaons or
nucleons. But even for pions, the $t^{*}=0$ approximation
merely leads to a slight overestimation (typically $<5\%$) of the
strong FSI effect \cite{led05}
and,
it doesn't influence the leading zero--distance
($r^{*}\ll |a|$) effect of the Coulomb FSI.

Then, for non--identical particles,
\begin{equation}
{\cal R}(p_{1},p_{2})\doteq
\sum_{S}\widetilde{\rho}_{S}
\langle |\psi_{-{\bf k}^{*}}^{S(+)}({\bf r}^{*})|^{2}
\rangle _{S},
\label{1}
\end{equation}
similar to the Fermi factor taking into account
the Coulomb FSI in $\beta-$decay.
For identical particles, the amplitude in Eq.~(\ref{1})
enters in a symmetrized form:
\begin{equation}
\label{sym}
\psi_{-{\bf k}^{*}}^{S(+)}({\bf r}^{*}) \rightarrow
[\psi_{-{\bf k}^{*}}^{S(+)}({\bf r}^{*})+(-1)^{S}
\psi_{{\bf k}^{*}}^{S(+)}({\bf r}^{*})]/\sqrt{2}.
\end{equation}
The averaging in Eq.~(\ref{1})
is done over the four--coordinates of the emitters
at a given total spin $S$ of the two--particles,
$\widetilde{\rho}_{S}$ is the corresponding population probability,
$\sum_{S}\widetilde{\rho}_{S} = 1$.
For unpolarized particles
with spins $s_{1}$ and $s_{2}$ the probability
$\widetilde{\rho}_{S}=(2S+1)/[(2s_{1}+1)(2s_{2}+1)]$.
Note that in the case of small $k^*$, we are interested in,
the short-range interaction is dominated by central forces and
s-waves so that, neglecting a weak spin dependence of the
Coulomb interaction, the spin dependence of the
two-particle amplitude enters only through the total spin $S$.

(v) The {\it two-particle approximation} in (i) and FSI separation
in the Bethe-Salpeter amplitudes of the elastic transitions
$1+2\to 1+2$ implies a long FSI time as compared with the
characteristic production time, i.e. the channel momentum $k^*$
much less than typical production momentum transfer of hundreds
MeV/$c$. In fact, the long-time FSI can be separated also in the
inelastic transitions, $1+2\to 3+4$, characterized by a slow
relative motion in both entrance and exit channels
\cite{lll97,led05}. The necessary
condition is an approximate equality of the sums of particle
masses in the channels $\alpha=1+2$ and $\beta=3+4$.
In the presence of such transitions the amplitudes in
Eqs.~(\ref{1}) and (\ref{sym}) should be
substituted by the solutions of the two-channel scattering
problem, i.e., formally, $S\to S,\alpha\alpha$ and $S,\beta\alpha$.
In practice, the particles $1,3$ and $2,4$ are members of the
same isomultiplets (as, e.g., in the transition
$\pi^- p\leftrightarrow \pi^0 n$)
so that one can put $\widetilde{\rho}_{S,\alpha\alpha}=
\widetilde{\rho}_{S,\beta\alpha}$ and assume the same
${\bf r}^{*}$-distributions in the channels $\alpha$ and $\beta$.

\section{Femtoscopy with identical particles}
For identical pions or kaons, the effect of the strong FSI is usually
small and the effect of the Coulomb FSI can be in first approximation
simply corrected for (see \cite{slape} and references therein).
The corrected correlation function is determined by the
QS symmetrization only (see Eq.~(\ref{sym}) and substitute the
non--symmetrized amplitude by the plane wave
${\rm e}^{i qx/2}$):
\begin{equation}
{\cal R}(p_{1},p_{2})= 1+\langle \cos(q\Delta x)\rangle .
\label{qscf}
\end{equation}
Its characteristic feature is
the presence of the interference maximum at small
components of the relative four-momentum $q$
with the width reflecting the inverse space-time extent
of the effective production region.

Note that the on-shell constraint
$q_0 = {\bf v}{\bf q}$
(${\bf v}={\bf P}/P_0$ is the pair velocity)
makes the $q$-dependence
of the correlation function essentially three--dimensional
\cite{kop74}
(particularly, in pair c.m. system,
$q\Delta x=-2{\bf k}^*{\bf r}^*$)
and thus makes impossible the unique Fourier reconstruction
of the space--time characteristics of the emission process.
However, within realistic models,
the directional and velocity
dependence of the correlation function
can be used to determine both
the duration of the emission and the form
of the emission region \cite{gkp71,KP72,kop74,pod89},
as well as - to reveal the details of the
production dynamics (such as collective flows;
see, e.g., \cite{PRA84,ham88,MAK88} and reviews
\cite{WH99,cso02}).
For this, the correlation functions can be analyzed
in terms of the {out} (x), {side} (y) and {longitudinal} (z)
components of the relative momentum vector
${\bf q}=\{q_x,q_y,q_z\}$ \cite{pod83,bdh94,pra95,csh95};
the {out} and {side} denote the transverse components
of the vector ${\bf q}$, the {out} direction is
parallel to the transverse component of the pair three--momentum.
The corresponding correlation widths are
usually parameterized in terms
of the Gaussian correlation radii $R_i$,
\begin{equation}
{\cal R}(p_{1},p_{2})=
1+\lambda\exp(-R_x^2q_x^2-R_y^2q_y^2-R_z^2q_z^2
-2 R_{xz}^2q_xq_z)
\label{osl}
\end{equation}
and their dependence on pair rapidity and transverse momentum
is studied.
The correlation strength parameter $\lambda$ represents
a fraction of the pairs of identical
pions or kaons emitted by independent short-lived sources.
The form of Eq.~(\ref{osl}) assumes azimuthal symmetry of the
production process \cite{WH99,pod83}. Generally, e.g.,
in case of the
correlation analysis with respect to the reaction plane, all three
cross terms $q_iq_j$ contribute \cite{wie_fi98}.

It is well known that particle correlations at high energies
usually measure only a small part of the space-time emission volume,
being only slightly sensitive to its increase related to the fast
longitudinal motion of particle sources. In fact,
due to limited source decay momenta
of few hundred MeV/$c$, the correlated particles with nearby velocities
are emitted by almost comoving sources and so - at nearby space--time
points.
The dynamical examples are sources-resonances,
colour strings or hydrodynamic expansion.
To substantially eliminate the effect of the longitudinal motion,
the correlations can be
analyzed in terms of the invariant variable
$Q = (-\widetilde{{q}}^2)^{1/2} = 2k^*$ and
the components of the  momentum difference in pair c.m. system
(${\bf q}^*\equiv {\bf Q}= 2{\bf k}^*$) or in
the longitudinally comoving system (LCMS) \cite{cso91}.
In LCMS each pair is emitted transverse to the reaction axis
so that the generalized relative momentum
$\widetilde{{\bf q}}$ coincides with ${\bf q}^*$
except for the component
$\widetilde{q}_x=\gamma_{t}q_x^*$,
where $\gamma_{t}$ is the LCMS Lorentz factor of the pair.

Particularly, in the case of one--dimensional boost invariant
expansion, the longitudinal correlation radius in the LCMS reads
\cite{MAK88} $R_z \approx (T/m_t)^{1/2}\tau$, where $T$ is the
freeze-out temperature, $\tau$ is the proper freeze-out time and
$m_t$ is the transverse particle mass.
In this model, the side radius measures the
transverse radius of the system while
the square of the out radius
gets an additional contribution $(p_t/m_t)^2\Delta\tau^2$
due to the finite emission duration $\Delta\tau$.
The additional transverse expansion leads to a slight
modification of the $p_t$--dependence of the longitudinal radius and -
to a noticeable decrease of the side radius and the spatial
part of the out radius with $p_t$.
Since the freeze-out temperature and the transverse flow determine
also the shapes of the $m_t$-spectra,
the simultaneous analysis of correlations and single particle
spectra for various particle species allows to disentangle
all the freeze-out characteristics
(see a review \cite{WH99}). It appears that with the increasing
energy of heavy ion collisions from AGS and SPS
up to the highest energies at RHIC,
the correlation data show rather weak energy dependence
\cite{adl04,ada05} and point to the kinetic
freeze-out temperature somewhat below the pion mass,
a strong transverse flow (with the mean transverse flow velocity
at RHIC exceeding half the velocity of light),
a short evolution time of 8-10 fm/$c$ and a very short
emission duration of about 2-3 fm/$c$ (see, e.g.,
a recent review \cite{lis05}).
The short evolution and emission duration at RHIC are also supported
by the correlation analysis with respect to the reaction plane
\cite{ada05,ada04}.

The small time scales at RHIC were not expected in
traditional transport and hydrodynamic models
(see, e.g., \cite{sbd01,hk02}) and may
indicate an explosive character of particle production
\cite{cso94,dp02} or, large parton cross sections \cite{lk02}
and early hadronization \cite{hum03} as in the
corresponding successful
multi-phase and hadron transport models.
In view of a successful hydrodynamical description of
the elliptic flow at RHIC (see, however, \cite{bor05})
the failure of hydrodynamics to
explain the femtoscopy data is often considered as a
puzzle.
However, it is still possible that the overestimation
of the outward and longitudinal radii will be cured
in the full three-dimensional
hydrodynamic calculations with the modified initial
conditions (earlier thermalization time or
initial transverse flow \cite{kr03,dp05}),
tuned equation of state with the account for
chemical non-equilibrium after hadronization
\cite{kr03,dp05,rap02,ht02}
and a better modeling of the freeze-out process (including
continuous emission during the hydrodynamic stage
\cite{ghk96,sah02} and
coupling to the hadron cascade \cite{bd00,tls01}).
This hope is supported by a better
description of the correlation radii
in a recent three-dimensional hydrodynamic
model \cite{ham05}
and by a successful description of the
SPS and RHIC data in a number
of hydro-motivated parametrizations (see a brief review
\cite{flo05}).
The success of these simple parametrizations
can be caused by the formation of
particle spectra and correlation radii
during an early stage of the kinetic freeze-out
(see, e.g., \cite{tls01,lk02,hum03}).
It was argued \cite{sah02} that such an early spectra formation
can be related with the fact that the solution of the
non-relativistic Boltzmann equation at spherically symmetric
initial conditions exactly coincides with free streaming
(see, e.g., \cite{ccl98}).
Indeed, numerical solutions of relativistic Boltzmann equation
at anisotropic initial conditions showed only a slight deviation
from free streaming
even in the case of a large number of rescatterings
\cite{almps05}.

\section{Femtoscopy with unlike particles}
The complicated dynamics of particle production,
including resonance decays and particle rescatterings,
leads to essentially non--Gaussian tail of the
$r^*$--distribution.
Therefore, due to different $r^*$--sensitivity
of the QS, strong and Coulomb FSI effects,
one has to be careful when analyzing the correlation functions
in terms of simple models.
Thus, the QS and strong FSI effects are influenced by the
$r^*$--tail
mainly through the correlation strength parameter $\lambda$ while,
the shape of the Coulomb FSI is sensitive to the distances
as large as
the pair Bohr radius $|a|$ (hundreds fm for the pairs
containing pions).
These problems can be at least partially overcome with the help of
transport code simulations accounting for the dynamical evolution
of the emission process and providing the phase space information
required to calculate the QS and FSI effects on the correlation
function.

Thus, in a preliminary analysis of the NA49 correlation data from
central $Pb+Pb$ 158 AGeV collisions \cite{led02},
the transport RQMD v.2.3 code was used.
To account for a possible
mismatch in $\langle r^*\rangle$, the correlation functions
were calculated with the space--time coordinates
of the emission points scaled by 0.7, 0.8 and 1.
The scale parameter was then fitted
using the quadratic interpolation.
The fits of the $\pi^+\pi^-$, $\pi^+ p$ and $\pi^- p$
correlation function indicate that RQMD
overestimates the distances $r^*$ by 10-20$\%$
thus indicating an underestimation of the collective
flow in this model.

Recently, there appeared data on $p\Lambda$ correlation functions
from $Au+Au$ experiments E985 at AGS \cite{lis01},
STAR at RHIC \cite{kis04} and
$Pb+Pb$ experiment NA49 at SPS CERN \cite{blu02}.
As the Coulomb FSI is absent
in $p\Lambda$ system, one avoids here the problem of its sensitivity
to the $r^*$--tail. Also, the absence of the Coulomb suppression
of small relative momenta makes this system more sensitive to the
radius parameters as compared with $pp$ correlations \cite{wan99}.
In spite of rather large statistical errors,
a significant enhancement is seen at low relative momentum,
consistent with the known singlet and triplet
$p\Lambda$ s--wave scattering lengths.
In fact, the fits using the analytical expression for the
correlation function
\cite{ll1} yield the correlation radii of 3-4 fm
in agreement with the radii obtained from $pp$ correlations in
the same experiments.
These radii are smaller than those obtained from two-pion
and two-kaon correlation functions at the same transverse
momenta and are in qualitative agreement with the
approximate $m_t$ scaling expected in the case of the
collective expansion.

\section{Accessing strong interaction}
In case of a poor knowledge of the strong interaction,
which is the case for meson--meson, meson--hyperon or
hyperon--hyperon systems,
it can be improved with the help of
correlation measurements.

In heavy ion collisions, the effective radius $r_0$ of the
emission region
can be considered much larger than the range of the
strong interaction potential.
The FSI contribution to the correlation function
is then independent of the actual
potential form \cite{ll1,gkll86}. At small $Q=2k^*$,
it is determined by the s-wave
scattering amplitudes $f^S(k^*)$ at a given total spin $S$
\cite{ll1}.
In case of $|f^S|>r_0$, this contribution is of the order of
$|f^S/r_0|^2$ and dominates over the effect of QS.
In the opposite case,
the sensitivity of the correlation function to the scattering
amplitude is determined by the linear term $f^S/r_0$.

The possibility of the correlation measurement
of the scattering amplitudes has been demonstrated \cite{led02}
in a preliminary analysis of the
NA49 $\pi^+\pi^-$ correlation data within the RQMD model.
The fitted strong interaction scale,
redefining the original s-wave scattering length of 0.23 fm,
appeared to be significantly
lower than unity: $0.63\pm 0.08$.
To a similar shift ($\sim 20\%$)
point also the recent BNL data on $K_{l4}$ decays \cite{pis01}.
These results are in agreement with the two--loop calculation
in the chiral perturbation theory with a standard value of the
quark condensate \cite{col00}.

Recently, also the singlet
$\Lambda\Lambda$ s--wave scattering length
has been estimated \cite{led02,blu02} based on the fits
of the NA49 data from $Pb+Pb$ collisions at
158 AGeV.
Though the fit results are not very restrictive, they
likely exclude
the possibility of a large singlet scattering length
comparable to that of $\sim$20 fm for the two--nucleon system.

The STAR experiment measured for the first time the
$p\bar{\Lambda}$ and $\bar{p}\Lambda$ correlation functions
and performed simultaneous fit of the correlation radius and
the spin-averaged s-wave scattering length.
The fitted imaginary part of the scattering length of $\sim 1$ fm
is in agreement with the $\bar{p}p$ results (thus pointing to
about the same $\bar{p}\Lambda$ and $\bar{p}p$ annihilation
cross sections) while the real part appears to be more negative
\cite{kis04}.

\section{Accessing relative space-time asymmetries}

The correlation function of two non--identical particles,
compared with the identical ones,
contains a principally new piece of information on the relative
space-time asymmetries in particle emission \cite{LLEN95}.
Since this information enters in the two-particle amplitude
$\psi_{-{\bf k}^{*}}^{S(+)}({\bf r}^{*})$
through the terms odd in
${\bf k}^*{\bf r}^*\equiv {\bf p}_1^*({\bf r}_1^*-{\bf r}_2^*)$,
it can be accessed studying the correlation functions
${\mathcal R}_{+i}$ and ${\mathcal R}_{-i}$
with positive and negative projection $k^*_i$ on
a given direction ${\bf i}$ or, - the
ratio ${\mathcal R}_{+i}/{\mathcal R}_{-i}$.
For example, ${\bf i}$ can be the direction of the pair
velocity or, any of the out (x), side (y), longitudinal (z)
directions. In LCMS,
$r^*_i=r_i$ except for
$r_x^*\equiv\Delta x^*=\gamma_{t}(\Delta x-v_{t}\Delta t)$,
where $\gamma_{t}$ and $v_t$
are the pair LCMS Lorentz factor and velocity.
One may see that the asymmetry in the out (x) direction
depends on both space and time asymmetries
$\langle\Delta x\rangle$ and $\langle\Delta t\rangle$.
In case of a dominant Coulomb FSI, the intercept of the correlation
function ratio is directly related with the asymmetry
$\langle r^*_i\rangle$ scaled by the Bohr radius
$a=(\mu z_{1}z_{2}e^{2})^{-1}$:
$
{\mathcal R}_{+i}/{\mathcal R}_{-i}\approx 1+
2\langle r_i^*\rangle /a.
$

A review of the simulation studies of the method sensitivity and
the experimental results can be found elsewhere \cite{led04}.
Here we only note that the out correlation asymmetries
between pions, kaons and protons observed
in heavy ion collisions at CERN
SPS and BNL RHIC
are in agreement with practically
charge independent meson production and a negative
$\langle\Delta x\rangle$
and/or positive $c\langle\Delta t\rangle$ on the level of
several fm (assuming $m_1<m_2$)
\cite{led02,blu02,ada03a,kis04,led04}.
In fact they are in quantitative agreement
with the RQMD transport model
as well as with the hydro-motivated blast wave parametrization,
both predicting the dominance of the spatial part of the
asymmetries generated by large transverse flows.
Particularly, the RQMD simulation
of central $Pb+Pb$ collisions
in conditions of the experiment NA49 at 158 AGeV
yields practically zero asymmetries for $\pi^+\pi^-$ system
while, for $\pi^\pm p$ systems,
$\langle\Delta x\rangle \doteq -5.2$ fm,
$\langle\Delta t\rangle \doteq 2.9$ fm/$c$,
$\langle\Delta x^*\rangle \doteq -8.5 fm$.
Besides, it predicts $\langle x\rangle$ increasing
with particle $p_t$ or
$v_t=p_t/m_t$, starting from zero due to kinematic reasons.
The asymmetry arises because of a faster increase with $v_t$
for heavier particle.
In fact, the hierarchy
$\langle x_\pi\rangle<\langle x_K\rangle<\langle x_p\rangle$
is a signal of a universal transversal collective
flow \cite{led02};
one should simply take into account that the mean thermal
velocity is smaller for heavier particle
and thus washes out the positive shift due to the
flow to a lesser extent.

These conclusions can be expressed in a simple analytical
form using a hydro-motivated parametrization.
Thus, assuming the longitudinal-boost invariance,
a linear non-relativistic
transversal flow velocity profile
$\beta_F=\beta_0 r_t/r_0$, the local thermal momentum distribution
characterized by the kinetic freeze-out temperature $T$
and the Gaussian density profile
$\exp(-r_t^2/(2r_0^2))$,
one confirms a faster rise of $\langle x\rangle$ with
$v_t$ for heavier particles \cite{led04,akk96}:
\begin{equation}
\label{x_hydro}
\langle x\rangle=
r_0{v_t\beta_0}/\left({\beta_0^2+T/m_t}\right).
\end{equation}
The size of the shift $\langle x\rangle$ and its mass
dependence are weakened when introducing the transverse
temperature gradient $\langle \Delta T/T\rangle_r$
like in the Buda-Lund model (this gradient just adds to
the denominator in Eq. (\ref{x_hydro})) \cite{cs02}.
The observation of large correlation asymmetries at RHIC
may thus point against the existence of a hot
emitting central zone and the corresponding gradient
$\langle \Delta T/T\rangle_r\sim 1$
indicated by the analysis of RHIC
data within the Buda-Lund model \cite{ccls04}.
Further quantitative studies are needed to clarify this
question.

\section{Conclusions}

Wealth of data on momentum correlations of various particle
species ($\pi^{\pm},K^{\pm 0},p^{\pm},\Lambda,\Xi$)
is available and gives unique space-time
information on production characteristics including
collective flows.
Rather direct evidence for a strong transverse flow in heavy ion
collisions at SPS and RHIC is coming from unlike particle
correlation asymmetries. Being sensitive to relative time delays
and collective flows,
the correlation asymmetries can be especially useful to study
the effects of phase transitions.
Weak energy dependence of correlation radii contradicts to
traditional hydrodynamic calculations which
overestimate longitudinal and, especially, outward radii
at RHIC. It remains to be clarified whether this can be cured
in full three-dimensional hydrodynamic calculations
with a more refined treatment of initial conditions,
equation of state and freeze-out process.
A number of succesful hydro motivated parametrizations give
useful hints in this direction.
The momentum correlations between specific particles
yield also a valuable information
on the strong interaction hardly accessible by other means.

\end{document}